\documentclass{article}
\usepackage{amsmath}
\usepackage{tabularx}
\usepackage{booktabs}

\usepackage{PRIMEarxiv}

\usepackage[utf8]{inputenc} 
\usepackage[T1]{fontenc}    
\usepackage{hyperref}       
\usepackage{url}            
\usepackage{booktabs}       
\usepackage{amsfonts}       
\usepackage{nicefrac}       
\usepackage{microtype}      
\usepackage{lipsum}
\usepackage{fancyhdr}       
\usepackage{graphicx}       
\graphicspath{{media/}}     

\title{
A Mutual Inclusion Mechanism for Precise Boundary Segmentation in Medical Images
\thanks{\textit{{Corresponding Author}}: 
\textbf{Junyi Xin, Teeradaj Racharak, Le-Minh Nguyen, Guanqun Sun}} 
}

\author{
  Yizhi Pan \\  
  School of Information Engineering \\ 
  Hangzhou Medical College \\
  Hangzhou, Zhejiang, 311399, China\\
  \texttt{panyizhi@hmc.edu.cn} \\
 \And
  Junyi Xin \\
  School of Information Engineering\\ 
  Hangzhou Medical College \\
  Hangzhou, Zhejiang, 311399, China\\
  \texttt{xinjunyi@hmc.edu.cn} \\
  \And
  Tianhua Yang \\
  School of Information Engineering\\ 
  Hangzhou Medical College \\
  Hangzhou, Zhejiang, 311399, China\\
  \texttt{yth@hmc.edu.cn} \\
  \And
   Teeradaj Racharak\\
  School of Information Science\\
  Japan Advanced Institute of Science and Technology \\
  Nomi, Ishikawa, 923-1292, Japan\\
  \texttt{racharak@jaist.ac.jp} \\
  \And
   Le-Minh Nguyen\\
  School of Information Science\\
  Japan Advanced Institute of Science and Technology \\
  Nomi, Ishikawa, 923-1292, Japan\\
  \texttt{nguyenml@jaist.ac.jp} \\
\And
  Guanqun Sun \\
  School of Information Engineering\\ 
  Hangzhou Medical College \\
  Hangzhou, Zhejiang, 311399, China\\
  School of Information Science\\
  Japan Advanced Institute of Science and Technology \\
  Nomi, Ishikawa, 923-1292, Japan\\
  \texttt{sun.guanqun@hmc.edu.cn} \\
}

\begin{document}
\maketitle

\begin{abstract}
In medical imaging, accurate image segmentation is crucial for quantifying diseases, assessing prognosis, and evaluating treatment outcomes. However, existing methods lack an in-depth integration of global and local features, failing to pay special attention to abnormal regions and boundary details in medical images. 
To this end, we present a novel deep learning-based approach, MIPC-Net, for precise boundary segmentation in medical images.
Our approach, inspired by radiologists' working patterns, features two distinct modules: (i) \textbf{Mutual Inclusion of Position and Channel Attention (MIPC) module}: To enhance the precision of boundary segmentation in medical images, we introduce the MIPC module, which enhances the focus on channel information when extracting position features and vice versa; (ii) \textbf{GL-MIPC-Residue}: To improve the restoration of medical images, we propose the GL-MIPC-Residue, a global residual connection that enhances the integration of the encoder and decoder by filtering out invalid information and restoring the most effective information lost during the feature extraction process. 
We evaluate the performance of the proposed model using metrics such as Dice coefficient (DSC) and Hausdorff Distance (HD) on three publicly accessible datasets: Synapse, ISIC2018-Task, and Segpc. Our ablation study shows that each module contributes to improving the quality of segmentation results. Furthermore, with the assistance of both modules, our approach outperforms state-of-the-art methods across all metrics on the benchmark datasets, notably achieving a 2.23mm reduction in HD on the Synapse dataset, strongly evidencing our model's enhanced capability for precise image boundary segmentation.
Codes will be available at https://github.com/SUN-1024/MIPC-Net.

\end{abstract}

\keywords{Mutual Inclusion \and Medical Image Segmentation \and Model Integration}

\section{Introduction}
Medical image segmentation plays a pivotal role in quantifying diseases, assessing prognosis, and evaluating treatment outcomes. It describes crucial observations in images, such as the degree, size, and location of lesions. However, manual segmentation by skilled professionals is both time-consuming and tedious \cite{fkd-med}. Therefore, with the advance of deep learning technologies, automatic medical image segmentation has attracted growing research interest.

Existing medical image segmentation methods usually follow the practice of combining Convolutional Neural Networks (CNNs) with Vision Transformer modules under the U-Net structure \cite{U-Net, FCN, ViT}.
For example, various U-Net variants have been proposed to improve medical image segmentation performance. ResUnet \cite{resunet}, Unet++ \cite{unet++}, and Unet3++ \cite{unet3++} introduced residual connections and complex skip connections, while Attention-Unet \cite{attention-unet} integrated attention mechanisms into the U-Net architecture. TransUNet \cite{transunet} and Swin-Unet \cite{swin-unet} incorporated Transformer and Swin-Transformer \cite{swin-Transformer} modules, respectively, to capture global information. 
However, medical image segmentation differs from generic image segmentation tasks. In medical image segmentation, data is characterized by small sample sizes and the need for precise boundary delineation. Unlike generic image segmentation models, which are required to cover all details of the image, medical image segmentation demands special attention to abnormal regions and boundary details in organ or pathological images. Therefore, local image features need to be combined with global features. To this end, attention mechanisms focusing on both channel and position information need to be introduced into the research.

In recent research, there has been a trend towards incorporating both channel and position attention mechanisms into models. SA-UNet \cite{sa-unet} and AA-TransUNet \cite{aa-transunet} incorporated spatial and channel attention, respectively, but lack comprehensive utilization of image features. TransUNet++ \cite{transunet++} and DS-TransUNet \cite{ds-TransUNet} integrated Transformers into skip connections but have limitations in overall architecture and feature integration. DA-TransUNet \cite{DA-Trans} merges position and channel attention but merely adapts a block from road segmentation, lacking tailored feature extraction for medical images.
These methods achieve improved performance over previous medical image segmentation models. However, they focus primarily on the overall segmentation overlap rather than specifically enhancing the boundary details of the segmentation results. Moreover, when extracting features from the perspective of channel and position, these models only focus on repeated feature extraction, potentially disrupting the original information without considering how to restore the boundary details of the image.

\begin{figure*}[t!]
\centering
\includegraphics[width=0.8\textwidth]{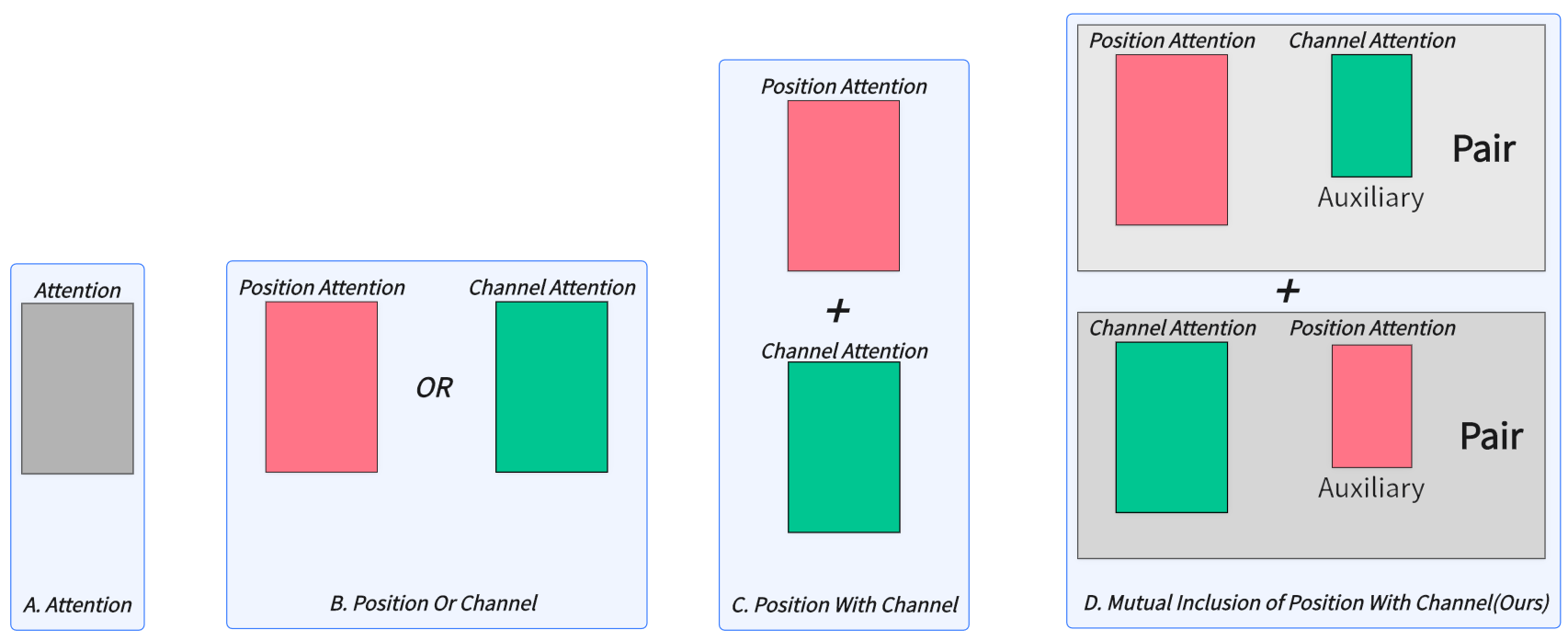}
\caption{\textbf{Comparison of attention mechanisms used in different medical image segmentation models: (a) only attention, (b) only channel or position attention, (c) integration of position and channel attention, and (d) Mutual nclusion of position and channel attention proposed in this work, which enhances the focus on channel information when extracting position features and vice versa"}}
\label{Fig_principle}
\end{figure*}

Inspired by radiologists' working patterns, this paper proposes a simple and effective mutual inclusion mechanism for medical image segmentation. Instead of simply stacking Transformer-related modules, we introduce the Mutual Inclusion of Position and Channel Attention (MIPC) module, which enhances the focus on channel information when extracting position features and vice versa. Figure \ref{Fig_principle} illustrates the superiority of our proposed mutual inclusion of position and channel attention compared to existing attention mechanisms.
We propose two pairs of channel and position combinations, each pair emphasizing either channel or position information while mutually including the other. This approach mimics the radiologists' working patterns, where mutual inclusion is practiced with varying emphasis. The experimental results demonstrate that this method effectively improves the model's ability to accurately segment image boundary.
Furthermore, we focus on the restoration of medical images by proposing the GL-MIPC-Skip-Connection. This connection introduces a Dual Attention mechanism to filter out invalid information while utilizing a global residual connection to restore the most effective information lost during the feature extraction process.


We evaluate our proposed methods on three publicly accessible datasets: the Synapse dataset \cite{Synapse}, the ISIC2018-Task dataset \cite{ISIC01, ISIC02}, and the Segpc dataset \cite{segpc}. In addition to the Dice coefficient (DSC) metrics, which deal with class imbalance problems, we adopt the Hausdorff Distance (HD) to analyze the quality of the segmentation results, as it is particularly convincing in evaluating boundary region segmentations. The results show that the proposed method achieves state-of-the-art performance on both DSC and HD metrics. Notably, there was a 2.23mm reduction over competing models in the HD metric on the benchmark Synapse dataset, strongly evidencing our model's enhanced capability for precise image boundary segmentation. This finding also indicates that medical image segmentation benefits from the mutual inclusion mechanism of position and channel attention.

The main contributions are as follows:
\begin{itemize}

    \item[1)]
    This paper proposes a novel model, MIPC-Net, which incorporates a Mutual Inclusion attention mechanism for position and channel information. This approach further enhances the precision of boundary segmentation in medical images.

    \item[2)]
    This paper introduces the GL-MIPC-Residue, a global residual connection that improves image restoration by enhancing the integration of the encoder and decoder.

    \item[3)]
    Experiments demonstrate that the proposed components achieve consistent performance improvements. Furthermore, our model achieves state-of-the-art performance across all metrics on the public Synapse \cite{Synapse}, ISIC2018-Task \cite{ISIC01, ISIC02}, and Segpc \cite{segpc} datasets.

\end{itemize}
The rest of this article is organized as follows. Section II reviews the related works of automatic medical image segmentation, and the description of our proposed MIPC-Net is given in Section III. Next, the comprehensive experiments and visualization analyses are conducted in Section IV. Finally, Section V makes a conclusion of the whole work.

\section{Related Work}

\subsection{Model Integration Of U-structure}
When constructing deep neural networks, properly utilizing residual learning and skip connections can enhance the model's learning capacity. Although the Unet model \cite{U-Net} enhances its functionality through three skip connections, it overlooks the optimization of these connections. Unet++ \cite{unet++} improves model performance by adding skip connections after a dense network architecture but fails to integrate other mechanisms for further optimization. Unet3++ \cite{unet3++} extends Unet++ by introducing hierarchical skip connections, enriching skip connections, and further enhancing the model's feature extraction capability, but only enriches skip connections without optimizing feature transmission during the process. DAResUNet \cite{daresunet} incorporates residual modules and DA-Blocks into skip connections but only optimizes the first-layer skip connection. DS-TransUNet \cite{ds-TransUNet} enhances skip connections by merging Transformer mechanisms into three skip connections but lacks further consideration of the overall structure. IB-TransUNet \cite{ib-TransUNet} integrates a multi-resolution fusion mechanism into skip connections but lacks comprehensive consideration of the overall model. DA-TransUNet \cite{DA-Trans} optimizes skip connections using image feature positions and channels, but overall model integration is insufficient. We not only optimize three skip connections but also further strengthen the overall integration of the model, achieving promising results in experiments.

\subsection{The Utilization Of Attention Modules.}
The attention mechanism is a crucial component that aids the model in focusing on target features and enhancing its performance. In recent years, there have been continuous advancements and iterations in the field of attention mechanisms.
In 2014, the concept of attention mechanisms was first introduced in the Bahdanau Attention paper \cite{bahdanau-attention}, initially applied in machine translation. 
In 2015, the introduction of attention mechanisms for image generation significantly enhanced the quality of the produced images \cite{gregor2015}.
Similarly, in 2015, the application of visual attention mechanisms to image description generation sparked substantial interest in the image captioning domain \cite{xu2015}.
In 2015, the introduction of various attention mechanism variants, such as global attention and local attention, marked a significant advancement \cite{luong2015}.
In 2017, the proposal of sub-attention mechanisms significantly propelled the evolution of attention mechanisms forward \cite{vaswani2017attention}.
In 2019, the employment of dual attention modules for scene segmentation, integrating both spatial and channel attention mechanisms, marked the pioneering introduction of dual attention mechanisms \cite{fu2019dual}.
The modular DAN (Dual Attention Network) framework, combining visual and textual attention, achieved significant outcomes in visual question-answering (VQA) tasks \cite{nam2017dual}.
The introduction of the Dual Attention Matching (DAM) module enhances high-level event inform   ation modeling over extended video durations, complemented by a global cross-check mechanism for precise localization of visible and audible events in videos \cite{wu2019dual}.
Furthermore, the application of dual attention mechanisms in medical image segmentation has shown promising results, but the strategies for optimizing feature extraction through position and channel attention mechanisms require further investigation \cite{shi2020clinically}.

In this paper, we propose a novel Mutually Inclusion of Position and Channel (MIPC) Block. By mutual including position and channel attention modules and incorporating the concept of residue, the segmentation performance of the model is significantly enhanced.






\section{Method}

In the following section, we introduce the MIPC-Net architecture, as depicted in Figure \ref{Fig_DDATrans-Unet}. We begin by providing an overview of the overall structure. Subsequently, we present its key components in the following sequence: Mutual Inclusion of Position and Channel (Section \ref{DDA-Block}), the encoder (Section \ref{Encoder}), GL-MIPC-Skip connections (Section \ref{skip}), and the decoder (Section \ref{Decoder}).


\subsection{Overview of MIPC-Net}
\begin{figure*}[t!]
\centering
\includegraphics[width=1\textwidth]{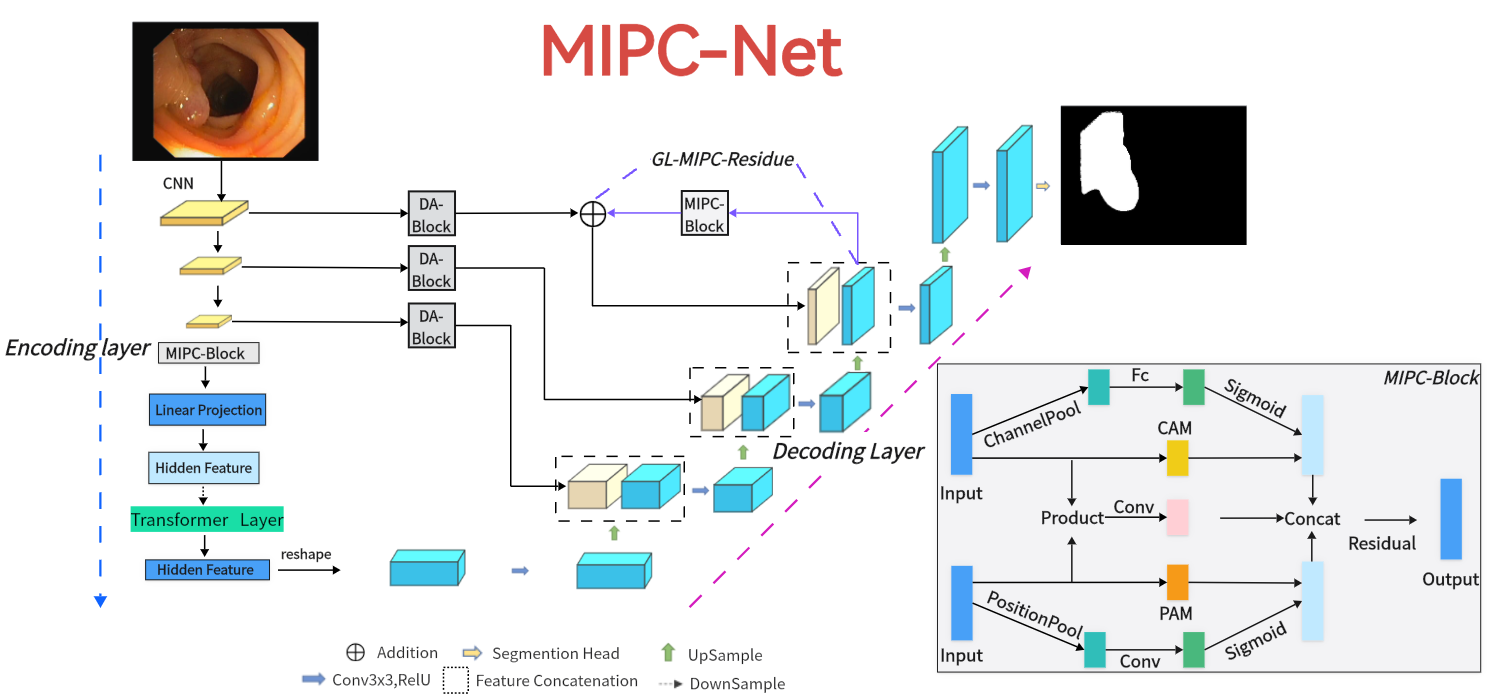}
\caption{\textbf{The illustration of the proposed MIPC-Net is depicted. For input medical images, they are fed into the encoder equipped with Transformer mechanisms and MIPC-Block. Subsequently, the features are restored to the original feature maps through the GL-MIPC-Skip-Connections and the decoder. This process yields the final image prediction results.}}
\label{Fig_DDATrans-Unet}
\end{figure*}

Figure \ref{Fig_DDATrans-Unet} illustrates the detailed configuration of our MIPC-Net model, which is a medical image segmentation model capable of capturing image-specific channel and position information and incorporates improved skip connections. 

Our model consists of three main components: the encoder, the decoder, and the GL-MIPC-Skip connections. 
Notably, the encoder integrates traditional convolutional neural network (CNN) and Transformer mechanisms, while using MIPC-Block to enhance the encoding capability (Section \ref{Encoder}). The decoder relies on deconvolution to restore the features to the original feature map size (Section \ref{Decoder}).
GL-MIPC-Skip-Connections employ DA-Block to purify the features of skip connection transmission. Furthermore, they use the GL-MIPC-Residue to further enhance the integrity of the encoder and decoder (Section \ref{skip}). 
MIPC-Net, comprised of three integral components, exhibits superior performance in image segmentation.

Given the constraints highlighted by traditional models, it is evident that while the conventional U-net architecture excels in capturing image features, it lacks effective methods for preserving and extracting global features.
On the other hand, Transformers exhibit remarkable proficiency in preserving and extracting global features through self-attention mechanisms \cite{transunet}. However, they are inherently limited to unidirectional positional attention, overlooking the utilization of image-special position and channel.
To address these limitations, we have integrated the Mutual Inclusion of Position and Channel Block (MIPC-Block) and leveraged GL-MIPC-Skip-Connections to enhance the integrity of the encoder and decoder, thereby improving medical image segmentation performance.
In medical image segmentation tasks, current models usually use attention mechanisms to enhance the segmentation capabilities of the model. For example: TransUNet uses ViT, and Swin-Unet uses Swin-Transformer. These approaches fail to adapt attention mechanisms to the specific features of the image, hence unable to extract deep image-related information.
To solve this problem, our proposed MIPC-Block enhances the segmentation capabilities of the model by leveraging image-specific features related to position and channel.
It effectively combines these two features in a mutually inclusive manner to extract deeper image-related features,  achieving subdivided extraction of image features and more fully mining features.

\begin{figure*}[t!]
\centering
\includegraphics[width=0.7\textwidth]{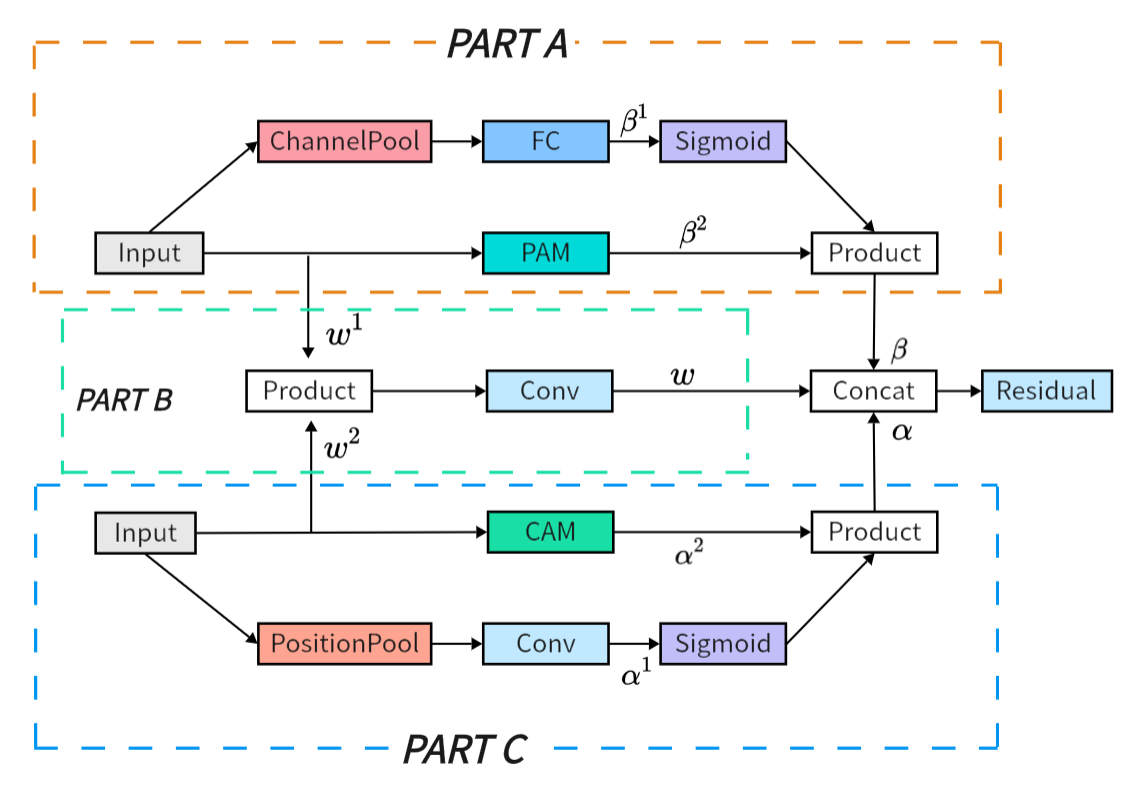}
\caption{\textbf{The proposed Position and Channel Mutual Inclusion Block (MIPC-Block) integrates positional, channel, and residual mechanisms. In Part A, attention is directed towards channels during the extraction of positional features, while in Part C, the reverse is applied.}}
\label{Fig_PCRA_principle}
\end{figure*}

As illustrated in Figure \ref{Fig_PCRA_principle}, the MIPC-Block architecture seamlessly integrates image-specific channel and positional features, enriched by the application of residual concepts. The amalgamation of channel and positional features empowers the MIPC-Block with profound insights into the image, surpassing the capabilities of conventional attention modules. 

The MIPC-Block architecture consists of three parts: PART A, PART B, and PART C. PART A and PART C serve as crucial feature extraction modules, ingeniously integrating both position and channel information of image features. The tight coupling of positional and channel information further enhances the feature extraction capability of the module. In Part A, our module undergoes a channel-wise average pooling layer (ChannelPool) to compress the feature map. Subsequently, it passes through fully connected layers to learn the correlations between different channels in the features. Following this, a sigmoid function is applied to constrain the values between 0 and 1, yielding channel correlations. Multiplying these correlations with the features obtained through the Position Attention Module (PAM) results in information where the position is the primary focus and channels act as auxiliary.
Conversely, in Part C, features are first subjected to MaxPool and AvgPool operations (PositionPool) along the spatial dimensions. The resulting features from these two pooling operations are concatenated, and through fully connected layers, correlations between different spatial dimensions in the features are learned. Similar to Part A, a sigmoid function constrains the values between 0 and 1. Multiplying these spatial correlations with the features obtained through the Channel Attention Module (CAM) produces information where channels are the main focus, and spatial dimensions serve as auxiliary.
Part B employs a residual approach to minimize the loss of valuable original information introduced by the convolution and attention modules.



\textbf{Part A (Position-Dominant Extraction with Channel)}: As illustrated in Figure \ref{Fig_PCRA_principle}, the extraction of channel information from the input features is facilitated by ChannelPool. Subsequently, a series of fully connected layers is employed to capture inter-channel correlations, yielding $\beta^{1}$. Concurrently, another set of input features undergoes processing by the Position Attention Module (PAM) to extract position information features, resulting in $\beta^{2}$. Following sigmoid processing of $\beta^{1}$, it is multiplied element by element with $\beta^{2}$ to obtain $\beta$. In contrast to Part C, where channel-wise modulation is utilized for distributing feature maps from the spatial module, this process generates feature maps with spatial and channel emphasis.

\begin{equation}
    \beta ^{1} = FC(ChannelPool( \text{Input} )),
\end{equation}

\begin{equation}
    \beta ^{2} = PAM( \text{Input} ),
\end{equation}

\begin{equation}
    \beta = Sigmoid( \beta ^{1} ) \cdot \beta ^{2},
\end{equation}

\textbf{PART B (Residual Part) }: As shown in the figure, Part A and Part B inputs undergo a convolutional operation to obtain $\omega^{1}$ and $\omega^{2}$, respectively. Subsequently, the two are element-wise multiplied and then passed through another convolutional layer to yield $\omega$. It extracts and refines features from both inputs, thereby refining the original features.
\begin{equation}
    \omega ^{1} = Conv(Part A's\quad Input),
\end{equation}
\begin{equation}
    \omega ^{2} = Conv(Part C's\quad Input),
\end{equation}
\begin{equation}
    \omega = Conv( \omega ^{1} \cdot \omega ^{2} )
\end{equation}

\textbf{PART C (Channel-Dominant Extraction with Position) }: As shown in Figure \ref{Fig_PCRA_principle}, the input features undergo PositionPool along the spatial dimension to effectively extract spatial information while eliminating noise and irrelevant details in the image. Subsequently, the feature maps are further processed by convolution to capture spatial correlations, resulting in $\alpha ^{1}$. Simultaneously, another set of input features is processed by the Channel Attention Module (CAM) to extract channel features, denoted as $\alpha^{2}$. The channel attention module is employed to extract detailed channel features from the image. After sigmoid processing of $\alpha ^{1}$, it is element-wise multiplied by $\alpha ^{2}$ to obtain the output $\alpha$. Unlike Part A, where the feature maps extracted by the spatial module are weighted by the channel attention module, effectively integrating image-specific spatial and channel features, generating feature maps with channel emphasis and spatial emphasis.

\begin{equation}
    \alpha ^{1} = Conv(PositionPool( \text{Input} )),
\end{equation}

\begin{equation}
    \alpha ^{2} = CAM( \text{Input} ),
\end{equation}

\begin{equation}
    \alpha = Sigmoid( \alpha ^{1} ) \cdot \alpha ^{2},
\end{equation}
Finally, the outputs of Parts A, B, and C are summed along the channel dimension, and then passed through a residual network (see Figure \ref{Residual}) to obtain the output.


\begin{equation}
    \text{Output} = \text{Residual}( \alpha + \beta + \omega ),
\end{equation}

\subsection{Mutual Inclusion of Position and Channel} 
\label{DDA-Block}

\begin{figure*}[t!]
\centering
\includegraphics[width=0.7\textwidth]{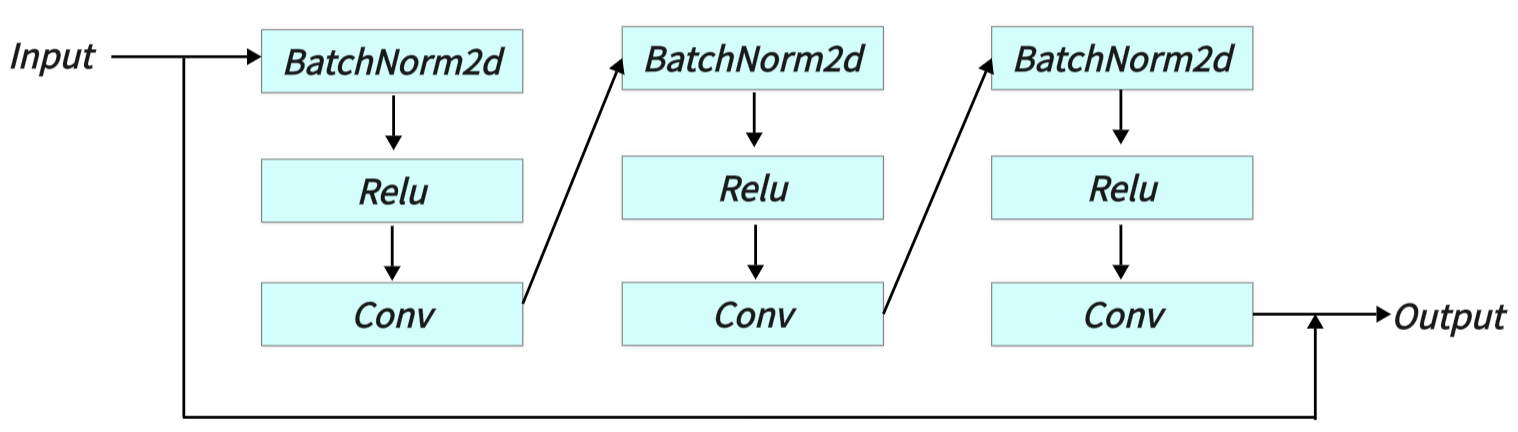}
\caption{\textbf{The specific structure of the last Residual module in MIPC-Block.}}
\label{Fig_MIPC_Residual}
\end{figure*}
\label{Residual}

The Mutual Inclusion of Position and Channel block (MIPC-Block) mutually includes the image features' position and channel, capturing deeper features associated with image features compared to standard attention modules.

\subsection{Encoder}
\label{Encoder}
As shown in Figure \ref{Fig_DDATrans-Unet}, the encoder consists of four key components: convolution blocks, MIPC-Block, an embedding layer, and transformer layers. 
It is particularly significant that the MIPC-Block is introduced just before the transformer layers. The purpose is to subject the convolutional features to specialized image processing, enhancing the transformer's feature extraction capabilities with respect to the image's content. 
The Transformer architecture excels at capturing global information. Integrating the MIPC-Block enhances its ability to maintain and extract global features specifically from images, enriching the Transformer's image processing capabilities.
This approach effectively combines image-specific channel and positional features with global features.

It begins with three U-Net convolutional blocks. Each block consists of a series of convolutions, normalization, and activation, designed to progressively refine input features, halve their size, and double their dimensions, thereby achieving efficient feature extraction. The MIPC-Block then purifies these features, emphasizing image-specific details for deeper analysis. An embedding layer adjusts feature dimensions for transformer layers, which address CNN limitations by capturing global information. Finally, the transformer's output is recombined and directed through skip connections to the decoder, ensuring comprehensive information retention and enhancing segmentation performance in a streamlined process.

By incorporating convolutional neural networks, transformer architecture, and Mutual Inclusion of Position and Channel, the encoder configuration ultimately attains robust feature extraction capabilities, resulting in synergistic strength.
\begin{figure*}[t!]
\centering
\includegraphics[width=0.7\textwidth]{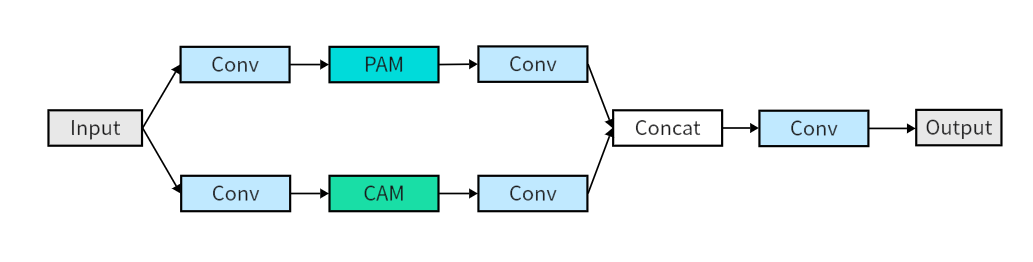}
\caption{\textbf{Architecture of Dual Attention Block (DA-Block).}}
\label{Fig_DA}
\end{figure*}
\subsection{GL-MIPC-Skip-Connections}
\label{skip}
Within the framework of the U-shaped encoder-decoder architecture, skip connections are utilized to alleviate semantic discrepancies between encoder and decoder components. However, the optimization of skip connections remains an area in need of improvement. Primarily, there exist challenges such as loss of feature fidelity during transmission and insufficient overall integrity between the encoder and decoder. 
To address these issues, we employed two strategies: purifying the features transmitted via skip connections and augmenting skip connections with global information. These approaches facilitate the decoder in accurately restoring the original feature map, thereby significantly enhancing the model's segmentation capabilities. Here we call the entire skip connection part GL-MIPC-SKip-Connections. It is divided into two parts: DA-SKip-Connections and GL-MIPC-Residue.
\subsubsection{DA-Skip Connections}

Analogous to the conventional U-structured models \cite{U-Net} \cite{daresunet}, our approach utilizes traditional skip connections to diminish the semantic disparity between the encoder and decoder. To further narrow this gap, we have incorporated Dual Attention Blocks (DA-Blocks) within all three skip connections, as illustrated in Figure \ref{Fig_DA}. This enhancement stems from our observation that features conveyed through skip connections frequently harbor redundancies, which DA-Blocks are adept at filtering out, thereby refining the feature transmission process.

The integration of Dual Attention Blocks (DA-Blocks) into skip connections empowers the model to meticulously refine features relayed from the encoder, through the lens of image-specific positional and channel-based considerations. This process facilitates the extraction of more pertinent information while minimizing redundancy. Such an enhancement not only bolsters the model's robustness but also significantly reduces the likelihood of overfitting, thereby contributing to superior performance and enhanced generalization capabilities.

\subsubsection{GL-MIPC-Residue}
Our distinction from other U-structured models lies in the sophisticated refinement of decoder features and their strategic incorporation into the skip connections, as illustrated in Figure \ref{Fig_DDATrans-Unet}. This approach is motivated by the realization that, although encoder features are extensively leveraged via skip connections, decoder features often remain underexploited. By purifying decoder features prior to their integration into skip connections—thereby enhancing the restoration process of the original feature map—we facilitate a more profound utilization of decoder features.

Purifying features within the decoder, after three stages of upsampling, using Mutual Inclusion of Position and Channel (MIPC-Blocks) — oriented specifically towards image-relevant channels and positions — significantly elevates the quality of information. Subsequent transmission of these enhanced features to the skip connections, followed by their integration into the decoder, ensures the comprehensive utilization of decoder features. This methodology effectively minimizes redundancy between encoder and decoder, enriches feature depth, mitigates overfitting risks, and augments the model’s capabilities in image segmentation and generalization.

\subsection{Decoder}
\label{Decoder}
As depicted in Figure \ref{Fig_DDATrans-Unet}, the diagram's right section represents the decoder. The decoder's fundamental task is to leverage features sourced from the encoder and those transmitted via skip connections. Through processes including upsampling, it endeavors to accurately reconstruct the original feature map.

The decoder architecture is structured around three pivotal elements: feature fusion, the segmentation head, and a series of three upsampling convolution blocks. Initially, feature fusion operates by amalgamating feature maps received through skip connections with current feature maps, thereby equipping the decoder to accurately reconstitute the original feature map. Subsequently, the segmentation head undertakes the task of adjusting the final output feature map back to its original dimensions. The final element comprises three upsampling convolution blocks, methodically increasing the size of the input feature map at each stage to adeptly reinstate the image's resolution.

Owing to the synergistic operation of these three components, the decoder showcases formidable decoding prowess. It adeptly harnesses features conveyed via skip connections as well as those derived from intermediate layers, enabling a proficient reconstruction of the original feature map.



\section{Experiment}

\subsection{Datasets}

In the Dataset section of our paper, we chose to conduct experiments on two distinct datasets: Synapse \cite{Synapse}, ISIC 2018 \cite{ISIC01, ISIC02} and Segpc \cite{segpc} for the following reasons:

Firstly, the Synapse dataset is among the most frequently utilized benchmark datasets in medical image segmentation, featuring segmentation tasks for eight different organs. This variety not only challenges but also demonstrates the generalization capabilities of our model across diverse anatomical structures.

Secondly, the selection encompasses both a 3D multi-class segmentation challenge (Synapse) and a 2D single-class segmentation task (ISIC 2018, Segpc). This combination allows us to evaluate our model's segmentation abilities from different perspectives, effectively showcasing its versatility and robustness in handling both complex three-dimensional data and simpler two-dimensional images.

This strategic choice of datasets underscores our commitment to validating the model's performance across a range of segmentation tasks, highlighting its potential for widespread application in medical image analysis.

\subsubsection{Synapse}
The Synapse dataset consists of 30 CT scan images of 8 abdominal organs.
organ. Including left kidney, right kidney, aorta, spleen, gallbladder, liver, spleen, pancreas and stomach,
liver, spleen, pancreas and stomach. A total of 3779 axial contrast-enhanced abdominal clinical CT images were obtained. In-plane resolution varies from 0.54 x 0.54 mm 2 to 0.98 x 0.98 mm2, while slice thickness ranges from 2.5 mm to 5.0 mm.

\subsubsection{ISIC-2018-Task}
The dataset used in the 2018 ISIC Challenge addresses the challenges of skin diseases. It comprises a total of 2512 images, with a file format of JPG. The images of lesions were obtained using various dermatoscopic techniques from different anatomical sites (excluding mucous membranes and nails). These images are sourced from historical samples of patients undergoing skin cancer screening at multiple institutions. Each lesion image contains only a primary lesion.

\subsubsection{Segpc}
This challenge targets robust segmentation of cells and is the first stage in building such tools for plasma cell cancers known as multiple myeloma (MM), a blood cancer. Provides images of stained colors normalized. The dataset contains a total of 298 images.

\subsection{Implementation Settings}

\subsubsection{Baselines}
In order to innovate in the field of medical image segmentation, we conducted benchmark testing of our proposed model against a series of well-regarded baselines, including U-net, UNet++, Residual U-Net, Att-UNet, TransUNet, and MultiResUNet. U-net has been a foundational model in the medical image segmentation domain \cite{U-Net}. UNet++ enriches the skip connections \cite{unet++}. Residual U-Net integrates a single residual module into the U-Net model \cite{resunet}, while MultiResUNet incorporates multiple residual modules \cite{multiresunet}. Att-UNet utilizes attention mechanisms to improve the weight of feature maps \cite{attention-unet}. Finally, TransUNet integrates the Transformer architecture, establishing a new benchmark in segmentation accuracy \cite{transunet}. Through comprehensive comparisons with these renowned baselines, our objective is to highlight the unique advantages and wide-ranging potential applications of our proposed model. Additionally, we benchmarked our model against advanced models. UCTransNet allocates attention modules in the traditional U-net model for skip connections \cite{uctransnet}, while MISSFormer moves attention module allocation into a Transformer module-based U-shaped structure \cite{missformer}. TransNorm integrates Transformer modules into the encoder and skips standard U-Net connections \cite{transnorm}. A novel Transformer module was designed, and a model named MT-UNet was constructed with it \cite{MT-UNet}. Swin-UNet further enhances segmentation by extensively applying Swin-transformer modules \cite{swin-unet}. DA-TransUNet enhances model segmentation capabilities by using image feature location contracts \cite{DA-Trans}. Through extensive comparisons with current state-of-the-art solutions, we aim to showcase its outstanding segmentation performance.

\subsubsection{Implementation Details}
We implemented MIPC-Net using the PyTorch framework and trained it on a single NVIDIA RTX 3090 GPU \cite{pytorch}.
The Transformer module we use employs the pre-trained model "R50-ViT". The input resolution and patch size set to 224x224 and 16, respectively. We trained the model using the SGD optimizer, setting the learning rate to 0.01, momentum of 0.9, and weight decay of 1e-4. The default batch size was set to 24. The loss function employed for dataset is defined as follows:
\begin{equation}
    \text{Loss} = \frac{1}{2} \times \text{Cross-Entropy Loss} + \frac{1}{2} 
    \times \text{DiceLoss}
\end{equation}

\subsubsection{Model Evaluation}
When evaluating the performance of MIPC-Net, we utilize a comprehensive set of metrics, including Intersection over Union (IoU), Dice Coefficient (DSC), and Hausdorff Distance (HD). These metrics are industry standards for computer vision and medical image segmentation and allow a multi-faceted assessment of a model's accuracy, precision, and robustness.

AC(Accuracy):
Accuracy is a widely used metric that assesses the overall correctness of a model's predictions. It calculates the proportion of correctly predicted samples over the total number of samples.Accuracy gives a general idea of how well the model is performing across all classes. 
\begin{equation}
    A C=\frac{T P+T N}{T P+T N+F P+F N}
\end{equation}


PR (Precision):
Precision focuses on the accuracy of the positive predictions made by the model. Precision is the ratio of correctly predicted positive observations to the total predicted positives.High precision indicates that the model is good at not misclassifying negative instances as positive.
\begin{equation}
    PR=\frac{T P}{T P+ F P}
\end{equation}


SP (Specificity):
Specificity measures the accuracy of negative predictions made by the model. Specificity is the ratio of correctly predicted negatives to the total predicted negatives. A high specificity suggests that the model is effective at correctly identifying true negatives.
\begin{equation}
    SP=\frac{T N}{T N+ F P}
\end{equation}



In summary, Accuracy provides an overall view of model performance, Precision emphasizes positive predictions' accuracy, and Specificity assesses the accuracy of negative predictions. 

IOU (Intersection over Union) is one of the commonly used indicators to evaluate the performance of computer vision tasks such as target detection, image segmentation and instance segmentation. It measures how much the model's predicted area overlaps with the actual target area, helping us understand the model's accuracy and precision. In image segmentation and instance segmentation tasks, IOU is used to evaluate the degree of overlap between predicted regions and ground-truth segmentation regions.

\begin{equation}I O U=\frac{T P}{F P+T P+F N}\label{iou}\end{equation}

The Dice coefficient (also known as Sørensen-Dice coefficient, F1-score, DSC) is a measure of model performance in image segmentation tasks and is particularly useful for dealing with class imbalance problems. It measures the degree of overlap between prediction results and ground-truth segmentation results, and is particularly effective when dealing with object segmentation with unclear boundaries. The Dice coefficient is commonly used in image segmentation tasks as a measure of the accuracy of the model in the target area.

\begin{equation}
          \operatorname{Dice}(P, T) = \frac{| P_{1} \cap T_{1} |}{| P_{1} | + | T_{1} |} \Leftrightarrow \text{Dice} = \frac{2 | T \cap P |}{| F | + | P|}
\label{ppv}\end{equation}

Hausdorff distance (HD) is a distance metric used to measure the similarity between two sets and is often used to evaluate the performance of models in image segmentation tasks. It is particularly useful in the field of medical image segmentation, where it can quantify the difference between predicted and true segmentations, and is particularly convincing in evaluating boundary region segmentations. The calculation of the Hausdorff distance captures the maximum difference between the true and predicted segmentation results.

\begin{equation}
H(A, B)=\max \left\{\max _{a \in A} \min _{b \in B}\|a-b\|, \max _{b \in B} \min _{ a \in A}\|b-a\|\right\}
\label{dice}\end{equation}

We use Dice and HD in the Synapse dataset, use the AC, PR, SP, Dice in the ISIC-2018-Task and Segpc datasets.

\subsection{Comparison to the State-of-the-Art Methods}
\subsubsection{Synapse}

To evaluate the performance of our proposed MIPC-Net model, we conducted extensive experiments on the widely-used Synapse multi-organ segmentation dataset \cite{Synapse}. We compared MIPC-Net with 12 state-of-the-art (SOTA) methods, including both CNN-based and transformer-based approaches, such as U-Net \cite{U-Net}, Res-Unet \cite{resunet}, TransUNet \cite{transunet}, U-Net++ \cite{unet++}, Att-Unet \cite{attention-unet}, TransNorm \cite{transnorm}, UCTransNet \cite{uctransnet}, MultiResUNet \cite{multiresunet}, Swin-Unet \cite{swin-unet}, MT-UNet \cite{MT-UNet}, and DA-TransUNet \cite{DA-Trans}. The experimental results are presented in Table \ref{table:Synapse}.

As shown in Table \ref{table:Synapse}, MIPC-Net achieves the highest average Dice Similarity Coefficient (DSC) of 80.00\% and the lowest average Hausdorff Distance (HD) of 19.32 mm among all the compared methods. This demonstrates the superior performance of MIPC-Net in both overall segmentation accuracy and boundary delineation precision. Compared to the popular transformer-based model TransUNet \cite{transunet}, MIPC-Net significantly improves the DSC by 2.52\% and reduces the HD by 12.37 mm, highlighting the effectiveness of our proposed mutual inclusion mechanism and global integration strategy.

Moreover, MIPC-Net consistently outperforms TransUNet in terms of DSC for all eight individual organs, with improvements ranging from 0.07\% to 4.12\%. Notably, MIPC-Net achieves substantial DSC improvements of 3.29\%, 3.35\%, 3.59\%, 4.12\%, and 3.93\% for the gallbladder, right kidney, pancreas, spleen, and stomach, respectively. These organs are known to be particularly challenging to segment due to their variable shapes, sizes, and locations, as well as their low contrast with surrounding tissues. The significant performance gains achieved by MIPC-Net demonstrate its strong capability in handling these difficult cases and accurately delineating organ boundaries.

\begin{figure*}[t!]
\centering
\includegraphics[width=0.7\textwidth]{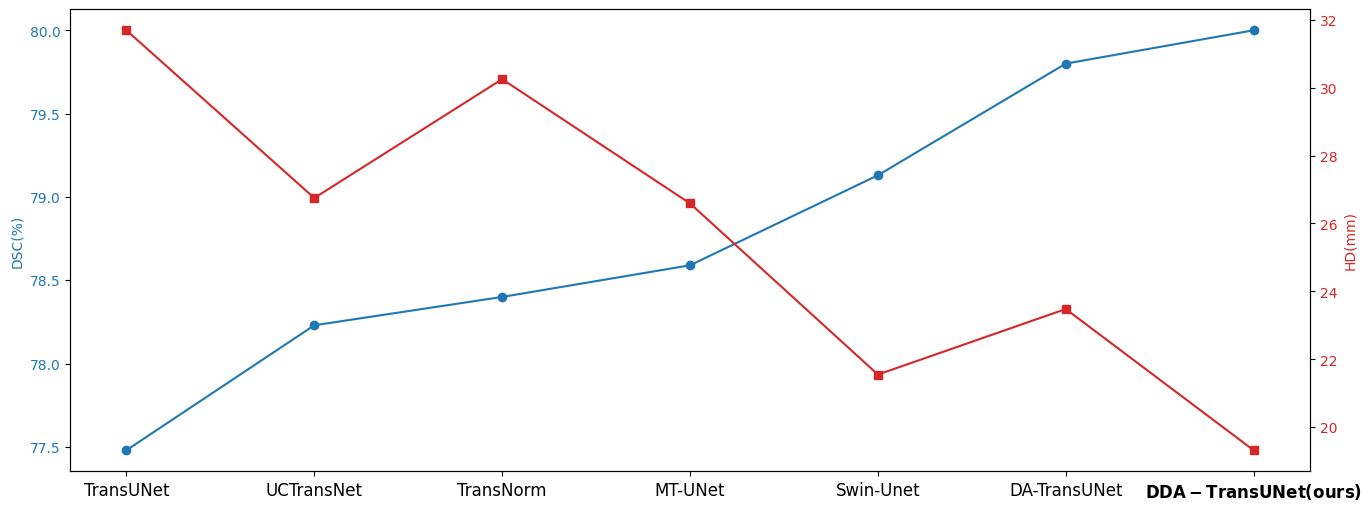}
\caption{\textbf{Line chart of DSC and HD values of several advanced models in the Synapse dataset}}
\label{Fig_line_chart}
\end{figure*}

Figure \ref{Fig_line_chart} provides a visual comparison of the DSC and HD values achieved by MIPC-Net and several other advanced models on the Synapse dataset. It is evident that MIPC-Net achieves the highest DSC and the lowest HD among all the compared models, further confirming its state-of-the-art performance in multi-organ segmentation.

To gain deeper insights into the boundary delineation performance of MIPC-Net, we also evaluated the HD metric for each individual organ, as shown in Table \ref{table: HD of Synapse}. MIPC-Net achieves the lowest HD for five out of eight organs, including the aorta, gallbladder, right kidney, pancreas, and stomach. Particularly, MIPC-Net significantly reduces the HD by 6.31 mm and 2.73 mm for the aorta compared to TransUNet and DA-TransUNet, respectively. These results highlight the superior boundary segmentation capability of MIPC-Net, which can be attributed to the effective integration of position and channel information through our proposed mutual inclusion mechanism.

It is worth noting that while MIPC-Net achieves state-of-the-art performance, its computational efficiency is comparable to that of TransUNet. The image segmentation time of MIPC-Net is 38.51 ms, only slightly higher than TransUNet's 33.58 ms. This indicates that the superior performance of MIPC-Net does not come at the cost of significantly increased computational overhead, making it a practical solution for real-world clinical applications.

\begin{figure*}[t!]
\centering
\includegraphics[width=0.7\linewidth]{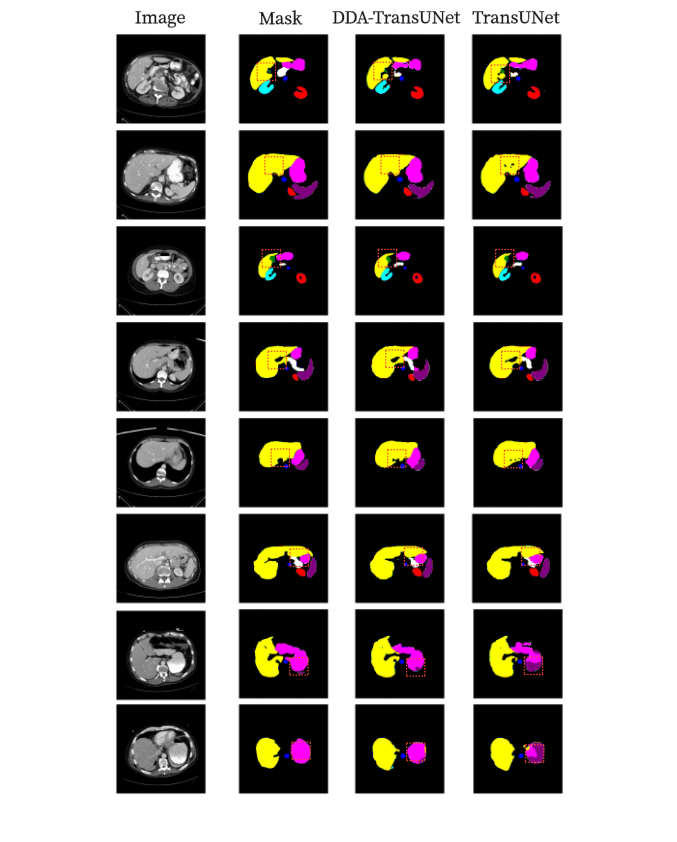}
\caption{\textbf{Segmentation results of TransUNet and MIPC-Net on the Synapse dataset.}}
\label{Segmentation of Synapse}
\end{figure*}

Figure \ref{Segmentation of Synapse} presents a qualitative comparison of the segmentation results produced by TransUNet and MIPC-Net on the Synapse dataset. The regions highlighted by orange borders clearly demonstrate that MIPC-Net generates more accurate and precise segmentations compared to TransUNet, especially in challenging areas such as organ boundaries and small structures. The visual results further validate the effectiveness of our proposed approach in capturing fine-grained details and producing high-quality segmentation masks.

\begin{table*}[!htbp]
\caption{\textbf{The experimental results on the Synapse dataset include the average Dice Similarity Coefficient (DSC) and Hausdorff Distance (HD) for each organ, as well as the individual DSC for each organ.}}
\label{table:Synapse}
\centering
\resizebox{\textwidth}{!}{%
\begin{tabular}{lcc|l|cccccccc}
\toprule
& & \multicolumn{2}{c}{mDSC, mHD} & \multicolumn{7}{c}{DSC of a single organ} \\
Model & Year & DSC$\uparrow$ & HD$\downarrow$ & Aorta & Gallbladder & Kidney(L) & Kidney(R) & Liver & Pancreas & Spleen & Stomach \\
\midrule
U-net \cite{U-Net} & 2015 & 76.85\% & 39.70 & 89.07 & 69.72 & 77.77 & 68.6 & 93.43 & 53.98 & 86.67 & 75.58 \\
U-Net++ \cite{unet++} & 2018 & 76.91\% & 36.93 & 88.19 & 68.89 & 81.76 & 75.27 & 93.01 & 58.20 & 83.44 & 70.52 \\
Residual U-Net \cite{resunet} & 2018 & 76.95\% & 38.44 & 87.06 & 66.05 & 83.43 & 76.83 & 93.99 & 51.86 & 85.25 & 70.13 \\
Att-Unet \cite{attention-unet} & 2018 & 77.77\% & 36.02 &89.55 & 68.88 & 77.98 & 71.11 & 93.57 & 58.04 & 87.30 & 75.75 \\
MultiResUNet \cite{multiresunet} & 2020 & 77.42\% & 36.84 & 87.73 & 65.67 & 82.08 & 70.43 & 93.49 & 60.09 & 85.23 & 75.66 \\
TransUNet \cite{transunet} & 2021 & 77.48\% & 31.69 & 87.23 & 63.13 & 81.87 & 77.02 & 94.08 & 55.86 & 85.08 & 75.62 \\
UCTransNet \cite{uctransnet} & 2022 & 78.23\% & 26.75 & 84.25 & 64.65 & 82.35 & 77.65 & 94.36 & 58.18 & 84.74 & 79.66 \\
TransNorm \cite{transnorm} & 2022 & 78.40\% & 30.25 & 86.23 & 65.1 & 82.18 & 78.63 & 94.22 & 55.34 & 89.50 & 76.01 \\
MT-UNet \cite{MT-UNet} & 2022 & 78.59\% & 26.59 & 87.92 & 64.99 & 81.47 & 77.29 & 93.06 & 59.46 & 87.75 & 76.81 \\
swin-unet \cite{swin-unet} & 2022 & 79.13\% & 21.55 & 85.47 & 66.53 & 83.28 & 79.61 & 94.29 & 56.58 & 90.66 & 76.60 \\
DA-TransUNet \cite{DA-Trans} & 2023 & 79.80\% & 23.48 & 86.54 & 65.27 & 81.70 & 80.45 & 94.57 & 61.62 & 88.53 & 79.73 \\
\textbf{MIPC-Net(Ours)} & ~ & \textbf{80.00\%} & \textbf{19.32} & 87.30 & 66.43 & 83.24 & 80.37 & 94.48 & 59.45 & 89.20 & 79.55 \\
\bottomrule
\end{tabular}%
}
\end{table*}

\begin{table}[!ht]
\caption{\textbf{The Hausdorff Distance (HD) for each organ in the Synapse dataset experimental results.}}
\label{table: HD of Synapse}
\centering
\begin{scriptsize} 
\begin{tabular}{ccccccccc}
    \hline
        Model & Aorta & Gallbladder & Kidney(L) & Kidney(R) & Liver & Pancreas & Spleen & Stomach \\ \hline
        TransUNet & 14.94mm & 15.81mm & 59.92mm & 45.76mm & 37.86mm & 17.34mm & 43.33mm & 18.56mm \\ 
        swin-unet & 8.64mm & 27.98mm & 41.83mm & 34.00mm & 22.17mm & 12.43mm& \textbf{9.90mm}& 15.45mm\\ 
        DA-TransUNet & 11.37mm & 27.93mm & \textbf{30.76mm}& 48.93mm & \textbf{20.26mm}& 12.29mm & 12.91mm & 23.37mm \\ 
        \textbf{MIPC-Net(Ours)} & \textbf{8.63mm}& \textbf{15.74mm}& 41.65mm & \textbf{27.12mm}& 22.33mm & \textbf{11.58mm}& 12.09mm & \textbf{15.39mm}\\ \hline
\end{tabular}
\end{scriptsize}
\end{table}

\subsubsection{ISIC 2018-Task Dataset}
\begin{table}[!ht]
\label{ISIC-table}
\caption{\textbf{Experimental results on the ISIC2018-Task dataset}}
    \centering
    \begin{tabular}{ccccc}
    \hline
        Method  & AC & PR & SP & Dice \\ \hline
        U-Net \cite{U-Net}  & 0.9446 & 0.8746 & 0.9671 & 0.8674  \\ 
        Att-UNet \cite{attention-unet}  & 0.9516 & 0.9075 & 0.9766 & 0.8820  \\ 
        U-Net++ \cite{unet++}  & 0.9517 & 0.9067 & 0.9764 & 0.8822  \\ 
        MultiResUNet \cite{multiresunet}  & 0.9473 & 0.8765 & 0.9704 & 0.8694  \\ 
        Residual U-Net \cite{resunet}  & 0.9468 & 0.8753 & 0.9688 & 0.8689  \\ 
        TransUNet \cite{transunet}  & 0.9452 & 0.8823 & 0.9653 & 0.8499  \\ 
        UCTransNet \cite{uctransnet}  & 0.9546 & 0.9100 & 0.9770 & \textbf{0.8898} \\ 
        MISSFormer \cite{missformer}  & 0.9453 & 0.8964 & 0.9742 & 0.8657 \\ 
        \textbf{MIPC-Net(ours)} & \textbf{0.9560} & \textbf{0.9279} & \textbf{0.9831} & 0.8875  \\ \hline
    \end{tabular}
\end{table}
To further validate the generalizability of MIPC-Net, we conducted experiments on the ISIC 2018 dataset \cite{ISIC01, ISIC02} for skin lesion segmentation. This dataset presents unique challenges, such as varying lesion sizes, shapes, and color variations.

Table \ref{ISIC-table} compares MIPC-Net with several state-of-the-art models on the ISIC 2018 dataset. MIPC-Net achieves the highest Accuracy (AC) of 0.9560, Precision (PR) of 0.9279, and Specificity (SP) of 0.9831, demonstrating its superior performance in accurately segmenting skin lesions. Notably, MIPC-Net significantly outperforms the transformer-based model TransUNet, with improvements of 0.0108 in AC, 0.0453 in PR, 0.0178 in SP, and 0.0376 in Dice index. These improvements can be attributed to the effectiveness of our proposed mutual inclusion mechanism and global integration strategy in capturing both local and global contextual information.

Interestingly, while MIPC-Net achieves the highest AC, PR, and SP, its Dice index of 0.8875 is slightly lower than that of UCTransNet (0.8898). This suggests a potential trade-off between precision and recall, which could be further investigated in future work.
\begin{figure*}[t!]
\centering
\includegraphics[width=1\linewidth]{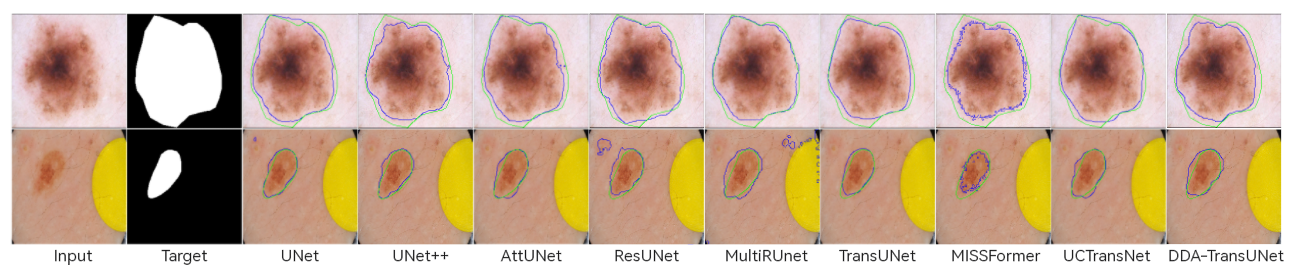}
\caption{\textbf{Segmentation results of TransUNet and MIPC-Net on the ISIC2018-Task dataset.}}
\label{Segmentation of ISIC}
\end{figure*}

Figure \ref{Segmentation of ISIC} qualitatively compares the segmentation results of TransUNet and MIPC-Net on the ISIC 2018 dataset. MIPC-Net generates more precise and accurate segmentations, especially in challenging cases with irregular lesion boundaries and low contrast. The visual results further validate the superiority of our approach in capturing fine-grained details and producing high-quality segmentation masks for skin lesions.
\subsubsection{Segpc Dataset}
\begin{table}[!ht]
\label{Segpc table}
\caption{\textbf{Experimental results on the Segpc dataset}}
    \centering
    \begin{tabular}{ccccc}
    \hline
        Method  & AC & PR & SP & Dice \\ \hline
        Residual U-Net \cite{resunet} & 0.9733 & 0.8917 & 0.9871 & 0.8479 \\
        MultiResUNet \cite{multiresunet} & 0.9753 & 0.8391 & 0.9834 & 0.8613 \\
        TransUNet \cite{transunet}  & 0.9671 & 0.8598 & 0.9882 & 0.8005 
        \\
        MISSFormer \cite{missformer}  & 0.9663 & 0.8152 & 0.9823 & 0.8082 \\ 
        DA-TransUNet \cite{DA-Trans} & 0.9713 & 0.8789 & 0.9845 & 0.8366 \\
        \textbf{MIPC-Net(ours)} & \textbf{0.9817}& \textbf{0.9079} & \textbf{0.9898} & \textbf{0.8675}  \\ \hline
    \end{tabular} 
\end{table}
We further assessed the performance of MIPC-Net on the Segpc dataset \cite{segpc} for cell segmentation in microscopy images. This dataset presents challenges such as overlapping cells, variable cell sizes and shapes, and low contrast between cells and background.

Table \ref{Segpc table} compares MIPC-Net with state-of-the-art models on the Segpc dataset. MIPC-Net consistently outperforms all compared methods, achieving the highest Accuracy (AC) of 0.9817, Precision (PR) of 0.9079, Specificity (SP) of 0.9898, and Dice index of 0.8675. Compared to TransUNet, MIPC-Net significantly improves performance across all metrics, with improvements of 0.0146 in AC, 0.0481 in PR, 0.0016 in SP, and 0.067 in Dice index. These substantial improvements demonstrate the effectiveness of our approach in accurately separating overlapping cells and dealing with low contrast.

Notably, MIPC-Net achieves a significantly higher Dice index (0.8675) compared to all other methods, indicating a good balance between precision and recall when segmenting cells, which is crucial for accurate cell analysis and quantification.

\begin{figure*}[t!]
\centering
\includegraphics[width=1\linewidth]{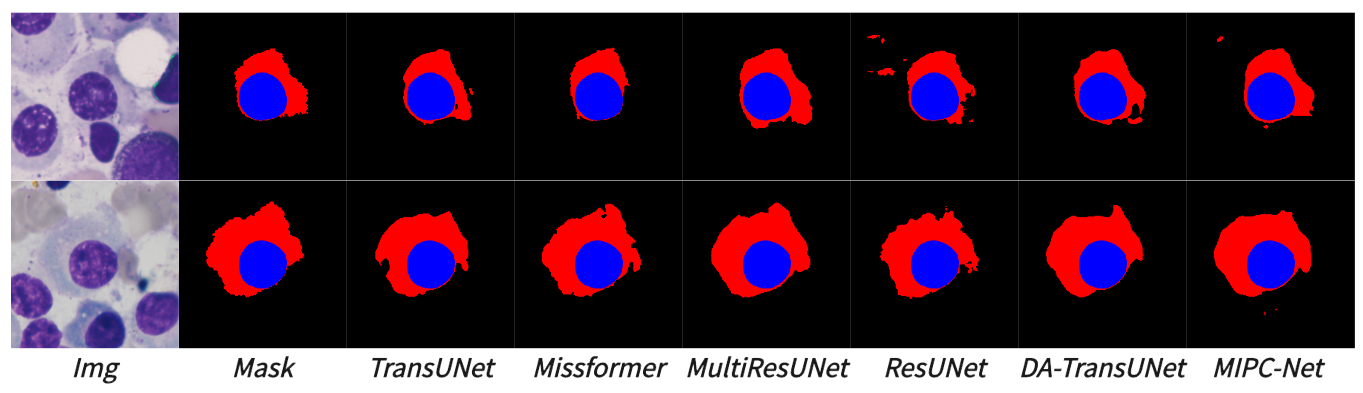}
\caption{\textbf{Segmentation results of TransUNet and MIPC-Net on the Segpc dataset.}}
\label{Segmentation of Segpc}
\end{figure*}

Figure \ref{Segmentation of Segpc} visually compares the segmentation results of TransUNet and MIPC-Net on the Segpc dataset. MIPC-Net generates more accurate and precise segmentations, successfully separating individual cells and capturing their fine boundaries, even in dense cell clusters.

The strong performance of MIPC-Net on the ISIC 2018 and Segpc datasets, along with its state-of-the-art results on the Synapse dataset, highlights the versatility and generalizability of our approach across different medical image segmentation tasks and modalities.


\subsection{Ablation Study}
To gain a deeper understanding of the effectiveness of the key components in our proposed MIPC-Net model, we conducted a comprehensive ablation study on the Synapse dataset. The study focused on three main aspects: the effects of mutual inclusion of position and channel, the impact of different configurations within the MIPC-Block, and the influence of the GL-MIPC-Residue in skip connections.
\subsubsection{The effects of Mutual Inclusion of Position and Channel}
\begin{table}[!htbp]
\caption{\textbf{Effects of Mutual Inclusion of Position and Channel}}
\label{table: Ablation OF MI}
\centering
\begin{tabular}{cccccc}
\hline
~ &  Mutual Inclusion & DSC$\uparrow$ & HD$\downarrow$ \\
\hline
PC-Net &  & 79.09 & 23.34 \\
MIPC-Net & $\surd$ & \textbf{80.00} & \textbf{19.32} \\
\hline
\end{tabular}
\end{table}
As shown in Table \ref{table: Ablation OF MI}, MIPC-Net, which incorporates the mutual inclusion mechanism, outperforms PC-Net by 0.91\% in terms of DSC and achieves a reduction of 4.02mm in HD. This improvement can be attributed to the effective integration of position and channel information through the mutual inclusion mechanism. By allowing the position and channel attention modules to interact and mutually guide each other, MIPC-Net is able to capture more comprehensive and discriminative features, leading to more accurate and precise segmentations. In contrast, simply using position and channel information independently, as in PC-Net, fails to fully exploit the potential synergies between these two types of information, resulting in suboptimal performance.
\subsubsection{The effects of how to mix MIPC-Block internal mechanisms}
%

\begin{table}[!htbp]
\caption{\textbf{Effects of how to mix MIPC-Block internal mechanisms}}
\label{table: Ablation01}
\centering
\begin{tabular}{ccccccc}
\hline
~ &Part.A Primary & Part.A Auxiliary &Part.C Primary & Part.A Auxiliary& DSC$\uparrow$ & HD$\downarrow$ \\
\hline
MIPC-Net&PAM&ChannelPool&CAM&PositionPool& \textbf{80.00} & \textbf{19.32} \\
MIPC-Net&PAM&ChannelPoll&PositionPool&CAM& 78.87 & 21.55 \\
MIPC-Net&ChannelPool&PAM&CAM&PositionPool& 79.10 & 26.38 \\
MIPC-Net&ChannelPool&PAM&PositionPool&CAM& 79.11 & 24.27 \\
\hline
\end{tabular}
\end{table}
Table \ref{table: Ablation01} presents the results of different configurations within the MIPC-Block. The optimal configuration, where position attention (PAM) is used as the primary focus and channel attention (ChannelPool) as the auxiliary focus in Part A, and channel attention (CAM) is used as the primary focus and position attention (PositionPool) as the auxiliary focus in Part C, achieves the best performance with a DSC of 80.00\% and an HD of 19.32mm. This suggests that a balance between position and channel attention is crucial for achieving the best segmentation results. By employing different primary attention modules in Part A and Part C, the MIPC-Block is able to capture complementary information from both position and channel perspectives, leading to more comprehensive feature extraction. Furthermore, the results demonstrate that using PAM and CAM as the primary attention modules consistently outperforms using ChannelPool and PositionPool as the primary modules, indicating that the self-attention mechanisms employed in PAM and CAM are more effective in capturing long-range dependencies and global contextual information.

\subsubsection{The effect of the GL-MIPC-Residue in skip connections}
\begin{table}[!htbp]
\caption{\textbf{Effects of the GL-MIPC-Residue in skip connections}}
\label{table:Ablation03}
\centering
\begin{tabular}{cccccccc}
\hline
 & \multicolumn{3}{c}{GL-MIPC-Residue} & & &\\

~ & 1st & 2nd & 3rd & DA-Skip-Connections& Encoder with MIPC& DSC$\uparrow$ & HD$\downarrow$ \\
\hline
MIPC-Net & &  & & $\surd $ & $\surd $ & 79.28 & 25.27 \\
MIPC-Net & $\surd $ & & & $\surd $ & $\surd $ & \textbf{80.00} & \textbf{19.32} \\
MIPC-Net & & $\surd $ & & $\surd $ & $\surd $ & 79.90 & 21.82 \\
MIPC-Net & & & $\surd $ & $\surd $ & $\surd $ & 78.64 & 27.78 \\
MIPC-Net & $\surd $ & $\surd $ & $\surd $ & $\surd $ & $\surd $ & 78.25 & 28.06 \\
MIPC-Net & &  & & & & 77.48 & 31.69 \\
\hline
\end{tabular}
\end{table}
Table \ref{table:Ablation03} shows the impact of the GL-MIPC-Residue module on the overall performance of MIPC-Net. Adding the GL-MIPC-Residue module to the first skip connection layer alone achieves the best performance, with a DSC of 80.00\% and an HD of 19.32mm, outperforming the baseline MIPC-Net without any GL-MIPC-Residue by 0.72\% in terms of DSC and reducing the HD by 5.95mm. This suggests that the GL-MIPC-Residue module is most effective when applied to the shallower skip connection layers, particularly the first layer, as it captures more low-level and spatial information crucial for accurate boundary delineation. The GL-MIPC-Residue module provides a direct path for the propagation of high-resolution spatial information from the encoder to the decoder, helping to preserve fine-grained details and improve localization accuracy. However, applying the GL-MIPC-Residue module to all skip connection layers leads to a significant performance drop, indicating that excessive use of the module can be counterproductive.

In conclusion, the ablation study demonstrates the importance of the mutual inclusion mechanism, the careful design of attention mechanisms within the MIPC-Block, and the strategic placement of the GL-MIPC-Residue module in skip connections. These components work together to capture comprehensive and discriminative features, leading to improved segmentation accuracy and precise boundary delineation in medical images.

\subsection{Discussion}
In this study, we found that the Mutual Inclusion of image-specific channels and positions can provide significant assistance for medical image segmentation tasks. The proposed MIPC-Block, based on the Mutual Inclusion mechanism, combined with GL-MIPC-Residue, further enhances the overall integration of the encoder and decoder. Our proposition has been validated through experiments on datasets, with the HD metric showing improvement to 2.23mm compared to competing models on the Synapse dataset, demonstrating strong boundary segmentation capabilities.

Analyzing the ablation experiments validates the effectiveness of our proposed MIPC Block and GI-MIPC-Residue. 
Firstly, according to the experimental results presented in Tables \ref{table: Ablation OF MI} and \ref{table: Ablation01}, we conclude that Mutual Inclusion of image feature positions and channels yields better performance compared to simple usage.
Additionally, as demonstrated by the results in Table \ref{table:Ablation03}, the GL-MIPC-Residue module enhances the overall integrity of the encoder-decoder. We conclude that reducing the loss of effective features is of paramount importance when deeply exploring features.

Despite these advantages, our model has some limitations.
Firstly, the introduction of MIPC-Block and DA-Blocks leads to an increase in computational complexity. This added cost may pose a barrier for real-time or resource-constrained applications.
Furthermore, this approach combines feature positions and channels attention with the Vision Transformer in a parallel manner, without achieving deep integration between them, indicating potential areas for further research and enhancement.

\section{Conclusion}
In conclusion, the proposed MIPC-Net represents a significant advancement in medical image segmentation, offering a powerful tool for precise boundary delineation. Inspired by radiologists' working patterns, our model integrates the Mutual Inclusion of Position and Channel Attention (MIPC) module and the GL-MIPC-Residue, a global residual connection, to effectively combine global and local features while focusing on abnormal regions and boundary details. The effectiveness of MIPC-Net is validated through extensive experiments on three publicly accessible datasets, outperforming state-of-the-art methods across all metrics and notably reducing the Hausdorff Distance by 2.23mm on the Synapse dataset. The mutual inclusion mechanism and the GL-MIPC-Residue contribute to the model's superior performance by allowing for a more comprehensive utilization of image features and enhancing the restoration of medical images. The improved precision in boundary segmentation has the potential to significantly impact clinical practice, leading to more accurate diagnosis, treatment planning, and ultimately better patient care. Future work may focus on extending the application of MIPC-Net to other medical imaging modalities, exploring its potential in tasks beyond segmentation, and incorporating domain-specific knowledge and multi-modal data to further enhance the model's performance and robustness.

\section*{Acknowledgments}
National Innovation and Entrepreneurship Training Program for College
Students (202113023001)

\bibliographystyle{unsrt}  
\bibliography{references}

\begin{thebibliography}{10}

\bibitem{fkd-med}
Guanqun Sun, Han Shu, Feihe Shao, Teeradaj Racharak, Weikun Kong, Yizhi Pan, Jingjing Dong, Shuang Wang, Le-Minh Nguyen, and Junyi Xin.
\newblock Fkd-med: Privacy-aware, communication-optimized medical image segmentation via federated learning and model lightweighting through knowledge distillation.
\newblock {\em IEEE Access}, 2024.

\bibitem{U-Net}
Olaf Ronneberger, Philipp Fischer, and Thomas Brox.
\newblock U-net: Convolutional networks for biomedical image segmentation.
\newblock In {\em Medical Image Computing and Computer-Assisted Intervention--MICCAI 2015: 18th International Conference, Munich, Germany, October 5-9, 2015, Proceedings, Part III 18}, pages 234--241. Springer, 2015.

\bibitem{FCN}
Jonathan Long, Evan Shelhamer, and Trevor Darrell.
\newblock Fully convolutional networks for semantic segmentation.
\newblock In {\em Proceedings of the IEEE conference on computer vision and pattern recognition}, pages 3431--3440, 2015.

\bibitem{ViT}
Alexey Dosovitskiy, Lucas Beyer, Alexander Kolesnikov, Dirk Weissenborn, Xiaohua Zhai, Thomas Unterthiner, Mostafa Dehghani, Matthias Minderer, Georg Heigold, Sylvain Gelly, et~al.
\newblock An image is worth 16x16 words: Transformers for image recognition at scale.
\newblock {\em arXiv preprint arXiv:2010.11929}, 2020.

\bibitem{resunet}
Foivos~I Diakogiannis, Fran{\c{c}}ois Waldner, Peter Caccetta, and Chen Wu.
\newblock Resunet-a: A deep learning framework for semantic segmentation of remotely sensed data.
\newblock {\em ISPRS Journal of Photogrammetry and Remote Sensing}, 162:94--114, 2020.

\bibitem{unet++}
Zongwei Zhou, Md~Mahfuzur Rahman~Siddiquee, Nima Tajbakhsh, and Jianming Liang.
\newblock Unet++: A nested u-net architecture for medical image segmentation.
\newblock In {\em Deep Learning in Medical Image Analysis and Multimodal Learning for Clinical Decision Support: 4th International Workshop, DLMIA 2018, and 8th International Workshop, ML-CDS 2018, Held in Conjunction with MICCAI 2018, Granada, Spain, September 20, 2018, Proceedings 4}, pages 3--11. Springer, 2018.

\bibitem{unet3++}
Huimin Huang, Lanfen Lin, Ruofeng Tong, Hongjie Hu, Qiaowei Zhang, Yutaro Iwamoto, Xianhua Han, Yen-Wei Chen, and Jian Wu.
\newblock Unet 3+: A full-scale connected unet for medical image segmentation.
\newblock In {\em ICASSP 2020-2020 IEEE international conference on acoustics, speech and signal processing (ICASSP)}, pages 1055--1059. IEEE, 2020.

\bibitem{attention-unet}
Ozan Oktay, Jo~Schlemper, Loic~Le Folgoc, Matthew Lee, Mattias Heinrich, Kazunari Misawa, Kensaku Mori, Steven McDonagh, Nils~Y Hammerla, Bernhard Kainz, et~al.
\newblock Attention u-net: Learning where to look for the pancreas.
\newblock {\em arXiv preprint arXiv:1804.03999}, 2018.

\bibitem{transunet}
Jieneng Chen, Yongyi Lu, Qihang Yu, Xiangde Luo, Ehsan Adeli, Yan Wang, Le~Lu, Alan~L Yuille, and Yuyin Zhou.
\newblock Transunet: Transformers make strong encoders for medical image segmentation.
\newblock {\em arXiv preprint arXiv:2102.04306}, 2021.

\bibitem{swin-unet}
Hu~Cao, Yueyue Wang, Joy Chen, Dongsheng Jiang, Xiaopeng Zhang, Qi~Tian, and Manning Wang.
\newblock Swin-unet: Unet-like pure transformer for medical image segmentation.
\newblock In {\em European conference on computer vision}, pages 205--218. Springer, 2022.

\bibitem{swin-Transformer}
Ze~Liu, Yutong Lin, Yue Cao, Han Hu, Yixuan Wei, Zheng Zhang, Stephen Lin, and Baining Guo.
\newblock Swin transformer: Hierarchical vision transformer using shifted windows.
\newblock In {\em Proceedings of the IEEE/CVF international conference on computer vision}, pages 10012--10022, 2021.

\bibitem{sa-unet}
Changlu Guo, M{\'a}rton Szemenyei, Yugen Yi, Wenle Wang, Buer Chen, and Changqi Fan.
\newblock Sa-unet: Spatial attention u-net for retinal vessel segmentation.
\newblock In {\em 2020 25th international conference on pattern recognition (ICPR)}, pages 1236--1242. IEEE, 2021.

\bibitem{aa-transunet}
Yimin Yang and Siamak Mehrkanoon.
\newblock Aa-transunet: Attention augmented transunet for nowcasting tasks.
\newblock In {\em 2022 International Joint Conference on Neural Networks (IJCNN)}, pages 01--08. IEEE, 2022.

\bibitem{transunet++}
Ali Jamali, Swalpa~Kumar Roy, Jonathan Li, and Pedram Ghamisi.
\newblock Transu-net++: Rethinking attention gated transu-net for deforestation mapping.
\newblock {\em International Journal of Applied Earth Observation and Geoinformation}, 120:103332, 2023.

\bibitem{ds-TransUNet}
Ailiang Lin, Bingzhi Chen, Jiayu Xu, Zheng Zhang, Guangming Lu, and David Zhang.
\newblock Ds-transunet: Dual swin transformer u-net for medical image segmentation.
\newblock {\em IEEE Transactions on Instrumentation and Measurement}, 71:1--15, 2022.

\bibitem{DA-Trans}
Guanqun Sun, Yizhi Pan, Weikun Kong, Zichang Xu, Jianhua Ma, Teeradaj Racharak, and Le-Minh Nguyen.
\newblock Da-transunet: Integrating spatial and channel dual attention with transformer u-net for medical image segmentation.
\newblock {\em arXiv preprint arXiv:2310.12570}, 2023.

\bibitem{Synapse}
Bennett Landman, Zhoubing Xu, Juan~Eugenio Igelsias, M~Styner, T~Langerak, and A~Klein.
\newblock Segmentation outside the cranial vault challenge.
\newblock In {\em MICCAI: Multi Atlas Labeling Beyond Cranial Vault-Workshop Challenge}, 2015.

\bibitem{ISIC01}
Noel Codella, Veronica Rotemberg, Philipp Tschandl, M~Emre Celebi, Stephen Dusza, David Gutman, Brian Helba, Aadi Kalloo, Konstantinos Liopyris, Michael Marchetti, et~al.
\newblock Skin lesion analysis toward melanoma detection 2018: A challenge hosted by the international skin imaging collaboration (isic).
\newblock {\em arXiv preprint arXiv:1902.03368}, 2019.

\bibitem{ISIC02}
Philipp Tschandl, Cliff Rosendahl, and Harald Kittler.
\newblock The ham10000 dataset, a large collection of multi-source dermatoscopic images of common pigmented skin lesions.
\newblock {\em Scientific data}, 5(1):1--9, 2018.

\bibitem{segpc}
Anubha Gupta, Ritu Gupta, Shiv Gehlot, and Shubham Goswami.
\newblock Segpc-2021: Segmentation of multiple myeloma plasma cells in microscopic images.
\newblock {\em IEEE Dataport}, 1(1):1, 2021.

\bibitem{daresunet}
Zhao Shi, Chongchang Miao, U~Joseph Schoepf, Rock~H Savage, Danielle~M Dargis, Chengwei Pan, Xue Chai, Xiu~Li Li, Shuang Xia, Xin Zhang, et~al.
\newblock A clinically applicable deep-learning model for detecting intracranial aneurysm in computed tomography angiography images.
\newblock {\em Nature communications}, 11(1):6090, 2020.

\bibitem{ib-TransUNet}
Guangju Li, Dehu Jin, Qi~Yu, and Meng Qi.
\newblock Ib-transunet: Combining information bottleneck and transformer for medical image segmentation.
\newblock {\em Journal of King Saud University-Computer and Information Sciences}, 35(3):249--258, 2023.

\bibitem{bahdanau-attention}
Dzmitry Bahdanau, Kyunghyun Cho, and Yoshua Bengio.
\newblock Neural machine translation by jointly learning to align and translate.
\newblock {\em arXiv preprint arXiv:1409.0473}, 2014.

\bibitem{gregor2015}
Karol Gregor, Ivo Danihelka, Alex Graves, Danilo Rezende, and Daan Wierstra.
\newblock Draw: A recurrent neural network for image generation.
\newblock In {\em International conference on machine learning}, pages 1462--1471. PMLR, 2015.

\bibitem{xu2015}
Kelvin Xu, Jimmy Ba, Ryan Kiros, Kyunghyun Cho, Aaron Courville, Ruslan Salakhudinov, Rich Zemel, and Yoshua Bengio.
\newblock Show, attend and tell: Neural image caption generation with visual attention.
\newblock In {\em International conference on machine learning}, pages 2048--2057. PMLR, 2015.

\bibitem{luong2015}
Minh-Thang Luong, Hieu Pham, and Christopher~D Manning.
\newblock Effective approaches to attention-based neural machine translation.
\newblock {\em arXiv preprint arXiv:1508.04025}, 2015.

\bibitem{vaswani2017attention}
Ashish Vaswani, Noam Shazeer, Niki Parmar, Jakob Uszkoreit, Llion Jones, Aidan~N Gomez, {\L}ukasz Kaiser, and Illia Polosukhin.
\newblock Attention is all you need.
\newblock {\em Advances in neural information processing systems}, 30, 2017.

\bibitem{fu2019dual}
Jun Fu, Jing Liu, Haijie Tian, Yong Li, Yongjun Bao, Zhiwei Fang, and Hanqing Lu.
\newblock Dual attention network for scene segmentation.
\newblock In {\em Proceedings of the IEEE/CVF conference on computer vision and pattern recognition}, pages 3146--3154, 2019.

\bibitem{nam2017dual}
Hyeonseob Nam, Jung-Woo Ha, and Jeonghee Kim.
\newblock Dual attention networks for multimodal reasoning and matching.
\newblock In {\em Proceedings of the IEEE conference on computer vision and pattern recognition}, pages 299--307, 2017.

\bibitem{wu2019dual}
Yu~Wu, Linchao Zhu, Yan Yan, and Yi~Yang.
\newblock Dual attention matching for audio-visual event localization.
\newblock In {\em Proceedings of the IEEE/CVF international conference on computer vision}, pages 6292--6300, 2019.

\bibitem{shi2020clinically}
Zhao Shi, Chongchang Miao, U~Joseph Schoepf, Rock~H Savage, Danielle~M Dargis, Chengwei Pan, Xue Chai, Xiu~Li Li, Shuang Xia, Xin Zhang, et~al.
\newblock A clinically applicable deep-learning model for detecting intracranial aneurysm in computed tomography angiography images.
\newblock {\em Nature communications}, 11(1):6090, 2020.

\bibitem{multiresunet}
Nabil Ibtehaz and M~Sohel Rahman.
\newblock Multiresunet: Rethinking the u-net architecture for multimodal biomedical image segmentation.
\newblock {\em Neural networks}, 121:74--87, 2020.

\bibitem{uctransnet}
Haonan Wang, Peng Cao, Jiaqi Wang, and Osmar~R Zaiane.
\newblock Uctransnet: rethinking the skip connections in u-net from a channel-wise perspective with transformer.
\newblock In {\em Proceedings of the AAAI conference on artificial intelligence}, volume~36, pages 2441--2449, 2022.

\bibitem{missformer}
Xiaohong Huang, Zhifang Deng, Dandan Li, Xueguang Yuan, and Ying Fu.
\newblock Missformer: An effective transformer for 2d medical image segmentation.
\newblock {\em IEEE Transactions on Medical Imaging}, 2022.

\bibitem{transnorm}
Reza Azad, Mohammad~T Al-Antary, Moein Heidari, and Dorit Merhof.
\newblock Transnorm: Transformer provides a strong spatial normalization mechanism for a deep segmentation model.
\newblock {\em IEEE Access}, 10:108205--108215, 2022.

\bibitem{MT-UNet}
Hongyi Wang, Shiao Xie, Lanfen Lin, Yutaro Iwamoto, Xian-Hua Han, Yen-Wei Chen, and Ruofeng Tong.
\newblock Mixed transformer u-net for medical image segmentation.
\newblock In {\em ICASSP 2022-2022 IEEE International Conference on Acoustics, Speech and Signal Processing (ICASSP)}, pages 2390--2394. IEEE, 2022.

\bibitem{pytorch}
Adam Paszke, Sam Gross, Francisco Massa, Adam Lerer, James Bradbury, Gregory Chanan, Trevor Killeen, Zeming Lin, Natalia Gimelshein, Luca Antiga, et~al.
\newblock Pytorch: An imperative style, high-performance deep learning library.
\newblock {\em Advances in neural information processing systems}, 32, 2019.

\end{thebibliography}

\end{document}